\documentclass[11pt]{article}

\usepackage[preprint]{acl}

\usepackage{times}
\usepackage{latexsym}
\usepackage[T1]{fontenc}
\usepackage[utf8]{inputenc}
\usepackage{microtype}
\usepackage{inconsolata}

\usepackage{graphicx}
\usepackage{booktabs}
\usepackage{amsmath,amssymb}
\usepackage{multirow}
\usepackage{xcolor}
\usepackage{placeins}

\newcommand{\pp}{\,pp}
\newcommand{\model}[1]{#1}
\newcommand{\code}[1]{\texttt{#1}}

\newcommand{\ph}{\code{\#~Your code here}}

\title{Solution or Skeleton? Distributional Alignment and\\
Mode Switching in Code Language Models}
\author{Shaojie Zhang\textsuperscript{1} \quad
  Yue Yu\textsuperscript{2} \quad
  Yu Zhang\textsuperscript{1}\thanks{Corresponding author: \texttt{yu.zhang.ust@gmail.com}.}\\
  \textsuperscript{1}Department of Computer Science and Engineering,\\
  Southern University of Science and Technology, Shenzhen, China\\
  \textsuperscript{2}Peng Cheng Laboratory, Shenzhen, China\\
  \texttt{12131109@sustech.edu.cn} \quad \texttt{yuy@pcl.ac.cn}}

\begin{document}
\maketitle

\begin{abstract}
A Python signature with a docstring and an empty body can open a solution file or an exercise skeleton.
We show that code language models resolve this ambiguity from the surrounding document and, when it resembles a skeleton, emit a \emph{deferral opener}---\ph{}, \code{pass}, or \code{raise NotImplementedError}---instead of code.
Across 24 open models, a neighbouring function can decide the choice: models copy its exact stub form rather than substitute another.
With nothing to copy, file-level text acts as a genre cue, and changing one word in a path raises deferral on every one of 23 comparable models.
In real GitHub files, models follow skeleton neighbours at the files' own rate, but the same cue raises false deferral on real code from $0.5\%$ to $7.1\%$.
Mechanistic analyses locate a late comment--code decision and a distributed reader--writer circuit.
Building on this structure, we introduce representation-level interventions that outperform coarse comment bans by $8.5$ and $5.5$\pp{} in two instruction-tuned HumanEval settings, while avoiding failures caused by suppressing useful comments.
The behaviour is specific to completion-style prompting and disappears under a chat template with free answers.
\end{abstract}

\section{Introduction}
\label{sec:intro}

\begin{figure*}[t]
\centering
\includegraphics[width=\textwidth]{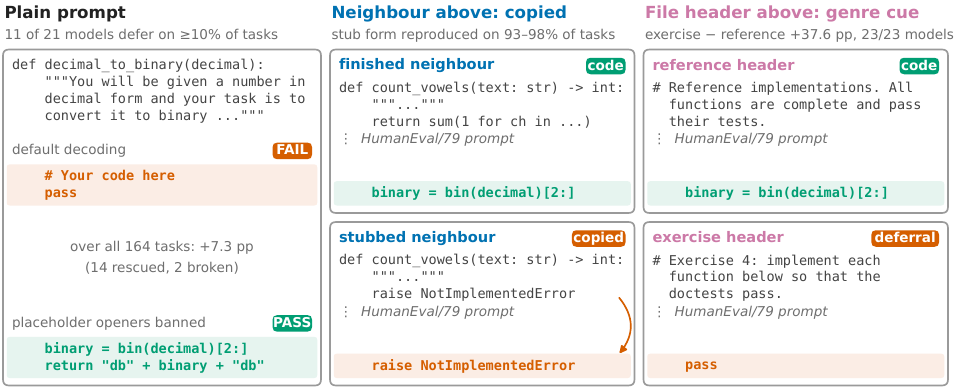}
\caption{\textbf{One task, five documents.}
Greedy first body line of \model{Coder-1.5B} on HumanEval/79.
Left: under the plain prompt the model writes \ph{} and \code{pass} and fails; banning placeholder openers at decode time yields a passing two-line solution.
Middle: one function above the task. After a finished neighbour the model writes code; after a stubbed one it copies the stub form.
Right: one header line above the task and nothing to copy (paraphrase pair 2 of 5; wrapped for display).
The single task is an illustration; the statistics above each column are medians over the 23 comparable models (Sections~\ref{sec:default}, \ref{sec:copy}, \ref{sec:genre}).}
\label{fig:overview}
\end{figure*}

Code language models (code LMs) are evaluated by completing a function from its signature and docstring~\cite{chen2021humaneval,liu2023evalplus}.
The same text also opens every exercise skeleton in a programming course, where the body is \ph{}, \code{\# TODO}, or \code{pass}.
A model that has seen both kinds of file must decide which one it is completing.
The question is not only academic: editor and repository-level completion feed the model the surrounding file~\cite{liu2024repobench,ding2023crosscodeeval}, and real files mix finished functions with stubs.

We show that code LMs make this decision from the surrounding document, and that when they decide ``skeleton'' they write a \emph{deferral opener} instead of code.
This matches the pre-training distribution, where a skeleton function is usually followed by another skeleton; Section~\ref{sec:natural} tests this on real files.
On HumanEval/79 (Figure~\ref{fig:overview}), \model{Coder-1.5B} writes \ph{} and fails; masking that opener yields a passing two-line solution.
This is not one model's quirk: under the plain HumanEval prompt, 11 of 21 open models from 7 families open at least 10\% of bodies with a placeholder comment (Section~\ref{sec:default}).
We study what in the document sets the decision, whether it matches real files, what it costs, and how it enters the network.

\paragraph{Findings.}
\begin{enumerate}
\item \textbf{Neighbouring functions are copied.} One function above the task moves deferral by at least 50\pp{} on every one of 21 comparable models, and the model reproduces the neighbour's stub form---\code{raise NotImplementedError} after \code{raise NotImplementedError}, \code{...} after \code{...}---almost never substituting another (Section~\ref{sec:copy}).
\item \textbf{File-level text sets a genre.} With nothing to copy, exercise-style headers raise deferral over reference-style headers on 23 of 23 models (median $+37.6$\pp{}), and one role word in a path, a course directory, or a statement that the file is unfinished is enough (Section~\ref{sec:genre}).
\item \textbf{The behaviour fits real files and misfires on them.} In real GitHub files, models defer after a skeleton neighbour about as often as the files do, and the same cue raises false deferrals on functions whose real body is code (Section~\ref{sec:natural}).
\item \textbf{Representation-level interventions help where useful comments make coarse bans brittle.} On instruction-tuned HumanEval, whitespace-gated steering exceeds a coarse comment ban by $+8.5$\pp{} on \model{Coder-1.5B-IT} and latent steering by $+5.5$\pp{} on \model{Qwen3-8B-IT}, under direct paired tests (Section~\ref{sec:discussion}). Probe and document gates provide targeted alternatives in complex-code and skeleton contexts.
\item \textbf{The two cues enter at different depths.} In an attention knock-out, the header's effect depends mainly on first-half attention and the neighbour's on second-half attention; the neighbour is copied through a specific but distributed set of heads that are mostly not induction heads (Section~\ref{sec:mech}).
\item \textbf{The cues also act in fill-in-the-middle (FIM) completion.} On four models and first lines only, a stubbed function \emph{below} the target raises deferral from $1.2\%$ to $27.4$--$49.4\%$ on \model{Coder-1.5B}, $62.8$--$67.1\%$ on \model{Coder-7B}, and $97.6$--$98.8\%$ on \model{Coder-3B}, and a stubbed function above still does (Section~\ref{sec:discussion}, Appendix~\ref{app:fim}).
\end{enumerate}
The effect is specific to completion-style prompts: under a chat template it shrinks when the code is prefilled and disappears when the model answers freely.
We approach mode switching through distribution analysis and mechanistic diagnosis, then use the resulting structure to target deferral without globally suppressing comments.

\section{Related Work}
\label{sec:related}

\paragraph{Comments and plans as scaffolds.}
Training on comment-dense code improves alignment between code and language~\cite{song2024codeneedscomments}.
At inference, MANGO asks for logical comments before code~\cite{chen2024mango}, and self-planning and structured chain-of-thought insert a plan first~\cite{jiang2024selfplanning,li2023scot,ma2023secot,lei2025planning}.
\citet{pan2025talking} show that comments describing a correct solution help weaker models.
\citet{ghoshal2026thinkless} find that a \code{\# Think step by step} prefix helps \model{Coder-1.5B} base and hurts its instruct variant.
These works study comments the user supplies.
We study the openers the model emits itself, and the document cues that make it emit them.

\paragraph{When context in code hurts.}
Incorrect documentation degrades code understanding~\cite{macke2024documentation}, TODO comments in a prompt can be reproduced rather than resolved~\cite{obrien2024todo}, and defective commented-out code raises completion defects~\cite{huang2026commenttraps}.
\citet{dinh2023buggy} show that a potential bug in the code prefix sharply lowers completion accuracy, and perturbing docstrings, names, or formatting changes code-generation results~\cite{wang2023recode}; prompt formatting alone moves LLM accuracy by large margins~\cite{sclar2024quantifying}.
These works show that context changes outputs.
We isolate one decision---solution or skeleton---and identify which document cues set it, whether it matches real files, and how each cue reaches it.
A benchmark prompt audit is a case of TODO copying: the LiveCodeBench prompt plants \code{\# YOUR CODE HERE}, and small models copy it (Appendix~\ref{app:lcb}).

\paragraph{Register, genre, and in-context copying.}
Register and genre condition word choice in human text~\cite{biber1988variation}; we ask whether code LMs infer the genre of a source file.
Induction heads copy earlier token patterns~\cite{olsson2022induction}; our stub-form test separates copying from substitution of another deferral, and head-level tests ask which heads carry the copy.

\paragraph{Localization and editing.}
We use the logit lens~\cite{nostalgebraist2020logitlens}, linear probes~\cite{alain2016understanding}, activation patching~\cite{meng2022locating,wang2023ioi,zhang2024patching}, and residual steering~\cite{turner2023activation,rimsky2024steering,zou2023representation}.
Our ban is a constrained-decoding restriction~\cite{scholak2021picard,geng2023grammar} used only as a probe.

\section{Setup}
\label{sec:setup}

\paragraph{Models.}
We use 24 open checkpoints from 13 families, including Qwen2.5-Coder (0.5B to 7B), Qwen3 (4B to 14B), DeepSeek-Coder, Gemma, Llama-3.1, StarCoder2, and others (full list in Appendix~\ref{app:models}).
First-line experiments add two more (26 checkpoints, 14 families).
Under the HumanEval prompt, StarCoder2-3B/7B and StableCode-3B skip the body at a token boundary, making their deferral rates incomparable; we exclude them from counts (leaving 21 or 23 comparable models).
Mechanistic analyses use \model{Qwen2.5-Coder-1.5B} (hereafter \model{Coder-1.5B}), \model{Coder-7B}, and \model{DeepSeek-Coder-1.3B-Base} (\model{DS-1.3B}).
\model{GrammarCoder-1.5B-Base} is included for its high placeholder rate despite its low pass@1 ($5.5\%$).

\paragraph{Task and metric.}
HumanEval ($n{=}164$) with EvalPlus base tests~\cite{chen2021humaneval,liu2023evalplus}, greedy decoding, raw completion.
A \textbf{placeholder} is a first-line comment deferring the work (regex: \code{your code}, \code{TODO}, \ldots) or an empty \code{\#}; a \textbf{stub} is \code{pass}, \code{...}, or \code{raise NotImplementedError}; a \textbf{deferral opener} is either.
\emph{Deferral} is the rate of deferral openers; \emph{placeholder rate} counts placeholders only.
First-line experiments generate until the first non-empty line.
Figures depend on the harness: for \model{Coder-1.5B}, $42.7\%$ and $42.1\%$ are pass@1 in the KV-cached split experiments and the HF \code{generate} experiments (Table~\ref{tab:after}); we compare only within a column.
Paired comparisons use two-sided exact McNemar tests ($^*p{<}0.05$).
For comparisons with several pairs, the task is the unit: we report task-bootstrap 95\% CIs and a sign-flip permutation test, Holm-corrected across models.
Every $\Delta$ is against the model's \code{none} cell in the same harness.

\paragraph{Document cues and FIM.}
A \textbf{neighbour} is one function placed above the task with a doctest docstring and one of seven bodies: finished code, one of five stub forms (\ph{} + \code{pass}, \code{...}, etc.), or no body.
We use three neighbours (\code{count\_vowels}, \code{merge\_intervals}, \code{is\_palindrome}).
A \textbf{file header} is one or two comment lines above the task: five exercise-style, five reference-style, and two neutral strings (Appendix~\ref{app:prompts}).
FIM runs use \model{Coder-1.5B}/\model{3B}/\model{7B} and StarCoder2-3B, vary the cue above or below the hole, and report first-line deferral only.

\paragraph{Decode-time and representation-level interventions.}
We compare lexical baselines with three interventions motivated by the late comment--code decision:
(1)~\code{ban} (coarse lexical mask) suppresses all comment tokens globally.
(2)~\code{ban\_placeholder} rewinds and suppresses only placeholder comments, leaving contentful comments intact.
(3)~\textbf{Probe-Gated Residual Editor}: a linear probe trained at statement starts (AUC $\approx 0.99$) detects deferral probability $p(\mathrm{comment})$; when $p \ge \tau$, it subtracts the separating direction ($h_\ell \leftarrow h_\ell - \alpha d_\ell$) without masking vocabulary tokens.
(4)~\textbf{Document Gate}: a linear classifier trained on prompt-end activations (AUC $0.90$--$0.999$) selectively gates representation editing to skeleton contexts.

\section{What the document tells the model}
\label{sec:frame}

\begin{figure*}[t]
\centering
\includegraphics[width=0.92\textwidth]{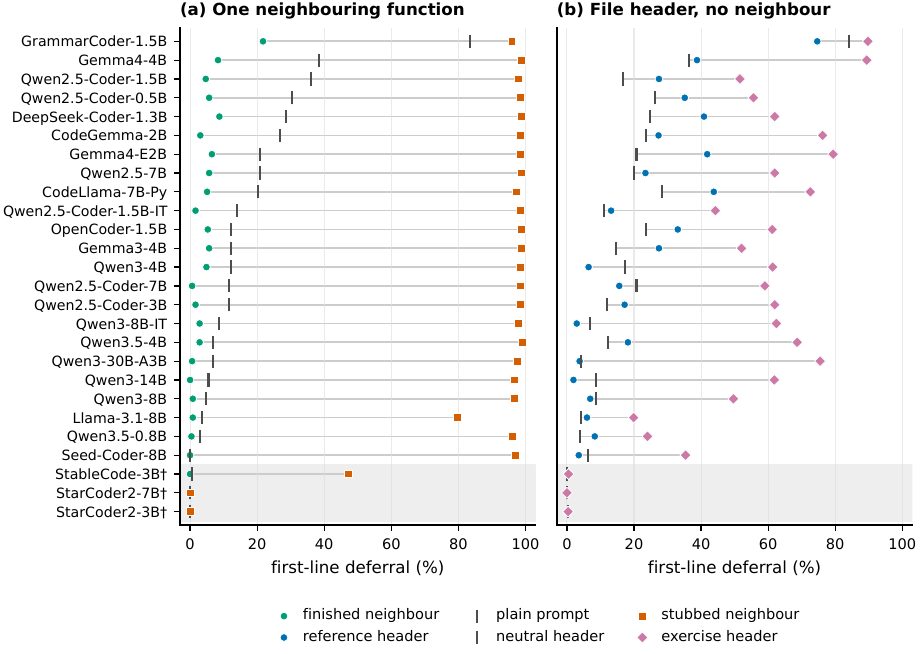}
\caption{\textbf{Deferral rate by document, 26 models.}
First-line deferral (placeholder, empty \code{\#}, or stub) on HumanEval. (a)~Plain prompt, and one finished or one stubbed (\ph{} + \code{pass}) neighbour, pooled over three neighbours. (b)~Exercise- and reference-style file headers (mean of five paraphrases each) and neutral headers (mean of two).
Models are sorted by the plain-prompt rate. Shaded, $\dagger$: models that skip the body at a token boundary under the plain prompt (excluded from counts).}
\label{fig:frame}
\end{figure*}

\subsection{One neighbour decides}
\label{sec:neighbour}
With one stubbed neighbour (\ph{} + \code{pass}), 20 of 24 models defer on ${\ge}96\%$ of tasks (Figure~\ref{fig:frame}).
This includes models that never defer under the plain prompt: \model{Seed-Coder-8B} goes from $0.0\%$ to $98.2\%$ and \model{Qwen3-14B} from $5.5\%$ to $97.6\%$.
With one finished neighbour, every model except GrammarCoder ($13.4\%$) is ${\le}7.9\%$.
The difference is at least 50\pp{} on all 21 comparable models (median $+95.7$\pp{}).

The effect saturates at one neighbour (tested up to four).
It also removes the ban gain: with two finished neighbours, \model{Coder-1.5B} and \model{Coder-7B} open no body with a placeholder and the ban changes pass@1 by $+0.6$ and $0.0$\pp{}.

\subsection{Copy, not substitution}
\label{sec:copy}
A stubbed neighbour could raise deferral in two ways.
The model could infer that it is in an exercise file and write whatever deferral it prefers, or it could copy the neighbour's body.
The stub forms separate the two: if the model infers a genre, a neighbour ending in \code{raise NotImplementedError} should sometimes yield \ph{}, the string models prefer under the plain prompt.

Table~\ref{tab:copy} shows that it does not.
Over 26 models (23 comparable), the target's first line is the neighbour's first body line on a median of $93$--$98\%$ of tasks for every stub form, and a different deferral form appears on a median of $0.0$--$0.2\%$.
The exception is GrammarCoder after a \code{...} neighbour (substituting \ph{} on $57.7\%$) and Qwen2.5-7B ($11.6\%$); no other model exceeds $6\%$.
Without a neighbour body to copy, deferral falls back near the plain-prompt rate (median $30.7\%$); \model{DS-1.3B} and CodeLlama copy even the absence, leaving the target body empty on ${\ge}93\%$ of tasks.
The neighbour effect is therefore in-context copying.

\begin{table}[t]
\centering
\small
\setlength{\tabcolsep}{3pt}
\resizebox{\columnwidth}{!}{%
\begin{tabular}{lrrrr}
\toprule
Neighbour body & Deferral & Copy & Other form & \model{Coder-1.5B} def. \\
\midrule
\ph{} + \code{pass} & 98.4 & 98.4 & 0.0 & 98.0 \\
\code{\# TODO: implement} + \code{pass} & 98.0 & 98.0 & 0.0 & 98.0 \\
\code{raise NotImplementedError} & 97.0 & 97.0 & 0.0 & 94.5 \\
\code{...} & 93.3 & 93.3 & 0.2 & 91.5 \\
\code{pass} & 96.5 & 96.5 & 0.0 & 85.6 \\
no body & 30.7 & -- & 30.7 & 32.1 \\
finished code & 3.0 & -- & 3.0 & 4.7 \\
\bottomrule
\end{tabular}%
}
\caption{\textbf{The neighbour's stub form is copied.}
Medians over 23 comparable models (\%, $n{=}3{\times}164$ per model: three neighbours); columns are medians taken separately.
Copy: the target's first line is the neighbour's first body line.
Other form: a deferral that is not the neighbour's.
Last column: \model{Coder-1.5B}; its plain-prompt deferral in the same run is $36.0\%$.}
\label{tab:copy}
\end{table}

\subsection{File-level text sets a genre}
\label{sec:genre}
Copying cannot explain deferral under the plain prompt, where no stub appears.
To test for a genre cue without anything to copy, we place a header above the task and no neighbour.
Exercise-style headers (\code{\# homework/hw3\_starter.py --- fill in the functions}, \code{\# Exercise 4: implement each function below so that the doctests pass.}, and three more) raise deferral over reference-style headers (\code{\# solutions/reference.py}, \code{\# answer\_key.py --- instructor solutions}, and three more) by a median of $+37.6$\pp{} (Table~\ref{tab:genre}).
The difference is positive on each of the 23 comparable models and significant on each after Holm correction (task-level test; sign test over models $p{=}2{\times}10^{-7}$), and 22 of them show it on all five paraphrase pairs (Qwen2.5-Coder-3B on four).
The effect is asymmetric: exercise headers raise deferral above a neutral header (\code{\# utils.py}) on 23/23 models (median $61.8\%$ vs.\ $16.8\%$), while reference headers do not push it below neutral (median $18.2\%$; below neutral on only 8/23).

The exercise headers also contain instructions (\code{fill in the functions}), so a model could be following an instruction rather than reading a genre.
Minimal pairs with no verb rule this out (Table~\ref{tab:genre}).
Changing one role word in a course path raises deferral by a median of $+16.6$\pp{}, and a neutral file name under a course path rather than a project path raises it by $+16.9$\pp{}; both are significant on each of the 23 models.
A sentence stating that the functions are not yet implemented raises it on 21 models; on CodeLlama-7B-Python and DeepSeek-Coder-1.3B the excess deferrals repeat the header sentence, and once these echoes are removed both go the other way.
The three cues hold on Qwen2.5-Coder-14B, DeepSeek-Coder-6.7B, and Qwen2.5-Coder-7B-Instruct (9/9 significant), and the exercise--reference gap holds on all 25 comparable models on MBPP+ (median $+30.8$\pp{}) and under sampling on three models ($T{=}0.2$ and $0.8$; Appendix~\ref{app:robust}).

\begin{table}[t]
\centering
\small
\setlength{\tabcolsep}{3pt}
\resizebox{\columnwidth}{!}{%
\begin{tabular}{lrrrr}
\toprule
Cue & $a$ & $b$ & $\Delta$ & Sig. \\
\midrule
Exercise ($a$) vs.\ reference ($b$) header & 61.8 & 18.2 & $+37.6$ & 23/0 \\
\midrule
\multicolumn{5}{l}{\emph{Minimal pairs ($a$ vs.\ $b$)}} \\
\code{\# hw3/starter.py} vs.\ \code{solution.py} & 55.1 & 33.8 & $+16.6$ & 23/0 \\
\code{\# hw3/utils.py} vs.\ \code{project/utils.py} & 35.8 & 18.9 & $+16.9$ & 23/0 \\
\code{\# \ldots{} not} vs.\ \code{fully implemented.} & 52.0 & 30.3 & $+22.0$ & 21/2 \\
\bottomrule
\end{tabular}%
}
\caption{\textbf{File text sets a genre.} First-line deferral (\%), medians over 23 models; no neighbour. Exercise and reference: mean over 5 paraphrases. Minimal pairs: medians over 5, 3, and 3 pairs; echoes of the header removed. Sig.: models significantly up/down (task-level test, Holm-corrected). Per-model rows in Table~\ref{tab:p2full} (Appendix~\ref{app:models}).}
\label{tab:genre}
\end{table}

Docstring style is a second, weaker cue.
Rewriting the doctest examples of the 76 HumanEval prompts that have them as prose (\code{For example, f(x) returns y.}) keeps the information and removes the \code{>>>} format; deferral rises on 15 of 23 models (falls on 4), significantly on \model{Coder-1.5B} ($+11.8$\pp{}$^*$), CodeGemma-2B ($+10.5$\pp{}$^*$), and Qwen2.5-7B ($+7.9$\pp{}$^*$).

\subsection{Not mainly difficulty}
\label{sec:difficulty}
A model might write a placeholder when it cannot solve the task.
We measure competence $c$ as pass@10 ($T{=}0.8$) under a finished neighbour, where placeholders are gone, and placeholder propensity $q$ as the logit of a placeholder opener under the plain prompt.
Within each of \model{Coder-1.5B}, \model{Coder-7B}, and \model{DS-1.3B}, $c$ explains 8--15\% of the variance in $q$; four surface features of the prompt (docstring length, number of \code{>>>} examples, number of \code{def}s, presence of an import) explain 33--47\%.
Of the tasks deferred under the plain prompt, $10/59$ (\model{Coder-1.5B}), $6/19$ (\model{Coder-7B}), and $9/47$ (\model{DS-1.3B}) are solved at least half the time once a finished neighbour is present.

\section{Real files}
\label{sec:natural}

To test hand-written cues against real code, we use the Python portion of the-stack-smol~\cite{kocetkov2022stack}, a 10,000-file GitHub sample.
None of its 23,617 documented functions opens with a placeholder comment (95\% upper bound $0.013\%$), against $34.8\%$ for \model{Coder-1.5B} under the plain HumanEval prompt.
Skeleton bodies that do occur are \code{pass} and \code{raise NotImplementedError} ($3.1\%$), and they cluster: a documented function is a skeleton $12.6\%$ of the time when its file contains another skeleton and $0.4\%$ otherwise.
So the default \ph{} is not explained by how often the string appears in public code; the clustering is.

We give every model the real file text up to each documented function's docstring: 1,430 functions with a skeleton function above them in a 6,000-character window, and 1,430 without.
With a skeleton above, the median model defers or leaves the body empty on $34.5\%$ of functions, against a real skeleton rate of $34.3\%$; without one, on $1.1\%$.
A median of $81.3\%$ of these deferrals copy the nearest skeleton's first line.
The same behaviour misfires: when the real body is code, a skeleton above raises false deferral from $0.5\%$ to $7.1\%$ (positive on all 21 comparable models; file-cluster bootstrap CIs exclude 0 on all), while a skeleton further down the same file, outside the prompt, leaves it at $0.9\%$.
Models therefore follow the cue at the rate real files do; on a benchmark or in an editor, where a body is always wanted, this document-appropriate behaviour is what costs accuracy.
Many of these files are likely in the models' training data: $40.6\%$ of model first lines equal the real one, so the calibration may partly be recall.

\section{The default prompt and its cost}
\label{sec:default}

\paragraph{How common.}
Under the plain HumanEval prompt, 11 of 21 pre-registered artefact-free models from 7 families open at least 10\% of bodies with a placeholder, from GrammarCoder-1.5B ($79.9\%$), Gemma4-4B ($38.4\%$), and Qwen2.5-Coder-1.5B ($34.8\%$) down to Qwen2.5-Coder-3B ($10.4\%$); Table~\ref{tab:p0} lists all models.
The rates are not tied to Python: on MultiPL-E Java~\cite{cassano2023multipl}, CodeGemma-2B opens $66.5\%$ of bodies with a \code{//} placeholder and Qwen3-8B-Instruct $41.1\%$ (Appendix~\ref{app:models}).

\paragraph{Where the ban gain lives.}
We split the effect of the decode-time ban by the first line of each model's own default completion.
On \model{Coder-1.5B}, the full ban gains $+8.5$\pp{}$^*$ ($18/4$ rescues/breaks).
Banning only placeholder openers gains $+7.3$\pp{}$^*$ ($14/2$); banning only other comments gains $+1.2$\pp{} ($3/1$, n.s.).
The gain sits on the 57 tasks whose default body opens with a placeholder ($+21.1$\pp{}$^*$); the 98 tasks that open in code move by $+3.1$\pp{} (n.s.).
The instruct checkpoint shows the same split: placeholder-only ban $+5.5$\pp{}$^*$ ($9/0$), other comments $+2.4$\pp{} (n.s.).
The gain is smaller and not significant on \model{Coder-7B} ($+1.8$\pp{}) and \model{DS-1.3B} ($+2.4$\pp{}).
Selectivity prevents collateral damage: on \model{Seed-Coder-8B} and \model{Coder-7B-IT}, coarse \code{ban} breaks 7 and 6 tasks (degrading \model{Coder-7B-IT} on HumanEval+ from $76.8\%$ to $76.2\%$), whereas \code{ban\_placeholder} incurs zero breaks on either ($+0.0$\pp{} and $+1.8$\pp{}; $+1.2$\pp{} on HumanEval+, $3/0$).
These bodies are not truncated: under the ban, completions get longer, and 65 of 95 default failures already contain wrong code after the opener (Appendix~\ref{app:errors}).

\paragraph{Across models.}
A high placeholder rate does not guarantee a significant gain.
Individually significant \code{ban\_placeholder} gains appear on GrammarCoder ($+21.3$\pp{}$^*$, $35/0$), \model{Coder-1.5B}, and \model{Coder-1.5B}-Instruct; Gemma4-4B defers on $38.4\%$ of tasks and gains $+1.8$\pp{} ($p{=}0.55$).
A pre-registered mixed model over 21 models (\code{pass $\sim$ ban + (1|model) + (1|task)}) gives $\beta{=}+0.22$ ($p{=}3{\times}10^{-5}$); dropping both \model{Coder-1.5B} checkpoints leaves $p{=}0.001$.
Dropping GrammarCoder as well, which was not pre-registered, leaves $\beta{=}+0.08$ ($p{=}0.17$).
The ban gain is individually significant only on GrammarCoder, \model{Coder-1.5B}, and \model{Coder-1.5B}-Instruct; for other models the confidence intervals include 0.

\begin{figure*}[t]
\centering
\includegraphics[width=0.92\textwidth]{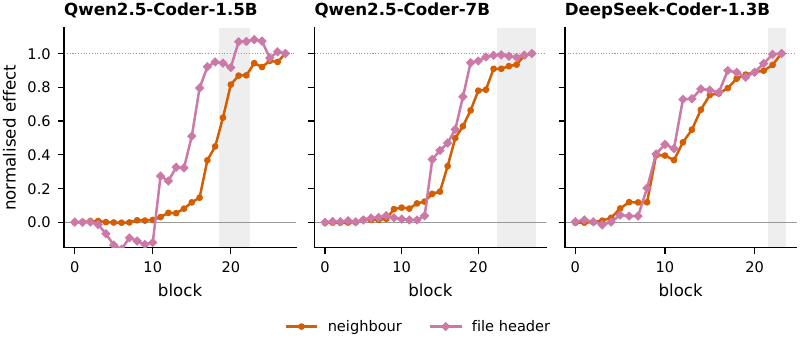}
\caption{\textbf{Where the document enters.} Normalised effect of patching each block's output at the decision positions (last prompt position and continuation). Neighbour: from the finished-neighbour run into the stubbed-neighbour run. File header: from the reference-header run into the exercise-header run (paraphrase pair $j$ cycles over tasks). 1 = the full gap in mean log-odds of the canonical first code line over \ph{}. Shaded: comment-surge layers from the logit lens.}
\label{fig:patch}
\end{figure*}

\section{Inside the model}
\label{sec:mech}

\paragraph{The decision is late.}
At the body start, the logit lens shows comment-opener mass near zero through mid layers and a surge in the last third (Figure~\ref{fig:lens} in the appendix): \model{Coder-1.5B} layers 19--22, \model{Coder-7B} 23--27, \model{DS-1.3B} 22--23, Qwen3-8B-Instruct 33--36.
On \model{Coder-1.5B}, comment mass is $0.03$ at layer 18 against $0.54$ for implementation tokens, the two meet at layer 20, and comment mass peaks at $0.77$.
A linear probe at statement starts separates comment from code at AUC $\approx 0.99$ in late layers.
The difference-in-means direction $d$ between comment and code states reads out mostly as comment-opener tokens (13 and 18 of its top 50 unembedding tokens on \model{Coder-1.5B} and \model{Coder-7B}, against about 4 by chance).
Adding $d$ to code-start states induces placeholders and failures (\model{Coder-1.5B}: 26 of 98 flips, $-9.8$\pp{}$^*$; \model{Coder-7B}: 68 of 131, $-19.5$\pp{}$^*$; matched random directions flip 20 and 4), and subtracting it acts like a soft ban (Appendix~\ref{app:steer}).
We use $d$ as a readout of the decision.

\begin{figure*}[t]
\centering
\includegraphics[width=0.92\textwidth]{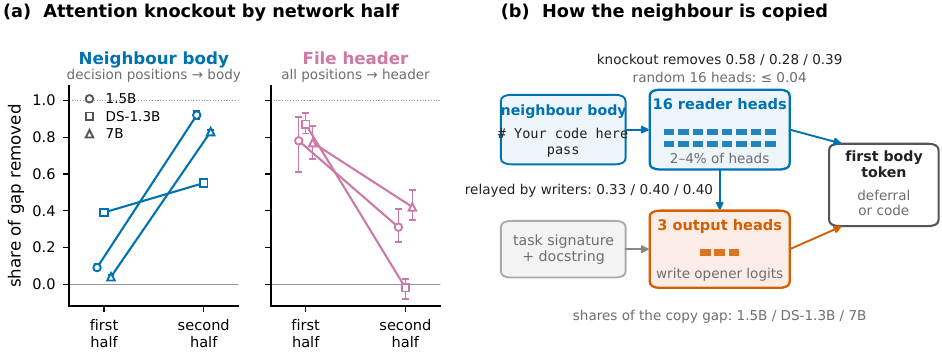}
\caption{\textbf{The two cues reach the decision differently}.
(a)~Share of each cue's gap in mean log-odds removed by knocking out attention in the first or second half of the network in both runs; HumanEval $n{=}164$, task-bootstrap 95\% CIs.
Neighbour: attention from the decision positions to the neighbour body. Header: attention from all positions to the header.
The cue $\times$ half interaction is 1.29, 1.06, and 1.14, with task-bootstrap intervals excluding 0.
(b)~How the neighbour is copied ($n{=}82$ held-out tasks). Knocking out the 16 heads with the most decision-to-body attention removes 0.58 / 0.28 / 0.39 of the copy gap (\model{Coder-1.5B} / \model{DS-1.3B} / \model{Coder-7B}); layer-matched random sets of 16 heads remove at most 0.04.
Freezing three output heads, which attend to the task text and write comment-opener logits, removes 0.33 / 0.40 / 0.40 of the readers' patching effect.}
\label{fig:access}
\end{figure*}

\paragraph{Where the document enters.}
We patch the output of each decoder block from the solution-cue run into the skeleton-cue run at the decision positions (last prompt position and continuation) and measure the log-odds of the canonical first code line over \ph{} (Figure~\ref{fig:patch}).
For the neighbour, patching has almost no effect before block ${\sim}16$; it reaches 50\% of the effect at block 19, 18, and 13 (\model{Coder-1.5B}, \model{Coder-7B}, \model{DS-1.3B}) and 90\% at block 23, 22, and 22, overlapping the comment surge.
For the header, it reaches 50\% at block 15, 17, and 12 and 90\% at block 17, 19, and 21: the header's information is present at the decision positions the same or a few blocks earlier.

\paragraph{The two cues enter at different depths.}
Block patching shows when information is present, not how it got there.
We therefore knock out attention edges in the first or the second half of the network, in both runs, and measure the share of each cue's gap removed (Figure~\ref{fig:access}a).
Knocking out attention by network half separates the two cues on all three models: the header effect loads on the first half ($0.78$, $0.87$, $0.77$) and the neighbour effect on the second ($0.92$, $0.55$, $0.83$).
The cue~$\times$~half interaction is $1.29$, $1.06$, and $1.14$, with task-bootstrap intervals excluding 0.
The cues also respond differently to editing: steering along a placeholder direction recovers the header gap ($1.22$, $2.78$, $1.41$ of it) but almost none of the copy gap ($0.07$, $0.23$, $0.01$), and the directions that carry the two cues overlap only partly ($|\cos|$ $0.14$--$0.38$).
On \model{DS-1.3B}, the neighbour effect is split across halves ($0.39$ / $0.55$).

\paragraph{The copy runs through specific but distributed heads.}
The 16 heads with the most attention from the decision position to the neighbour's body (2--4\% of all heads) remove $0.58$, $0.28$, and $0.39$ of the copy gap when knocked out; three layer-matched random sets of 16 heads remove at most $0.04$, and the top 8 alone remove only $0.19$--$0.26$ (Figure~\ref{fig:access}b).
Induction heads play a small role: of the 12 heads with the largest single-head patching effect, only 2, 1, and 1 (\model{Coder-1.5B}, \model{DS-1.3B}, \model{Coder-7B}) pass a prefix-matching test, and induction heads carry $13\%$, $7\%$, and $8\%$ of the positive single-head effect (Appendix~\ref{app:route}).
Part of the readers' effect is relayed: freezing three output heads, which attend to the task text and write comment-opener logits, removes $0.33$, $0.40$, and $0.40$ of the readers' patching effect.
Although the behaviour is verbatim copying (Section~\ref{sec:copy}), the circuit is a distributed reader set feeding a few writers, not the canonical prefix-matching induction circuit.

\section{Discussion}
\label{sec:discussion}

\paragraph{Completion mode and FIM.}
The behaviour belongs to editor completion, FIM, and base-model harnesses.
Under a chat template, prefilled code prompts retain stub effects ($37.0\%$ on \model{Coder-1.5B-IT}) while free answers show $0\%$ deferral (Appendix~\ref{app:robust}).
In FIM (Appendix~\ref{app:fim}), prefix cues survive ($74.4$--$98.6\%$ deferral), and a stubbed function \emph{below} the hole also raises deferral ($27$--$99\%$), with head-level patching effects correlating moderately ($r{\approx}0.54$).

\paragraph{A neighbour is not an evaluation fix.}
Across 505 default-deferral tasks (20 models), a finished neighbour rescues 78 tasks and placeholder ban rescues 67, but breaks 246 non-deferring tasks ($p{=}0.03$); rankings barely move ($\tau{=}0.87$).

\paragraph{Targeted gains from representation-level interventions.}
Table~\ref{tab:intervention_main} reports direct paired comparisons with the relevant lexical baseline.
The clearest gains occur on instruction-tuned models, where comments can carry useful reasoning.
On \model{Coder-1.5B-IT}, whitespace-gated steering improves over coarse \code{ban} by $8.5$\pp{} ($24/10$, $p{=}0.024$); on \model{Qwen3-8B-IT}, latent steering improves over it by $5.5$\pp{} ($10/1$, $p{=}0.012$).
The latter is significant relative to \code{ban}, although its $+5.5$\pp{} change from allow is marginal ($p{=}0.093$), because \code{ban} trades 13 rescues for 13 breaks.

\begin{table}[t]
\centering
\small
\setlength{\tabcolsep}{3.2pt}
\resizebox{\columnwidth}{!}{%
\begin{tabular}{@{}lllrr@{}}
\toprule
Setting & Lexical baseline & Intervention & $\Delta$ & $p$ \\
\midrule
1.5B-IT, HE & comment ban & WS-gated & $+8.5$ & .024 \\
8B-IT, HE & comment ban & latent steer & $+5.5$ & .012 \\
1.5B, BCB & comment ban & probe-gated & $+3.4$ & .30 \\
7B, BCB & comment ban & WS-gated & $+2.7$ & .455 \\
DS-1.3B, \code{...} & deferral ban & document-gated & $+7.9$ & $<.05$ \\
7B, ex.\ header & deferral ban & document-gated & $+10.4$ & $<.05$ \\
\bottomrule
\end{tabular}%
}
\caption{\textbf{Where representation-level interventions help.}
Direct paired pass@1 difference (percentage points) relative to the named lexical baseline.
HE: HumanEval; BCB: BigCodeBench-hard.
The BCB differences are not significant; document-gated gains are specific to the indicated model--shell settings.}
\label{tab:intervention_main}
\end{table}

BigCodeBench exposes a different boundary.
On \model{Coder-1.5B}, coarse \code{ban} changes pass@1 by $-2.0$\pp{}, while probe gating changes it by $+1.4$\pp{}; on \model{Coder-7B}, the corresponding changes are $+2.0$ and $+4.7$\pp{}.
The direct advantages in Table~\ref{tab:intervention_main} are not significant, so they show that statement-level gating can avoid the direction of coarse-ban damage, not a general BCB win.
Finally, document gating adds to a first-token deferral ban in two targeted shell settings ($+7.9$ and $+10.4$\pp{}), while the remaining shells show no reliable increment.
Appendix~\ref{app:steer} reports the full model--benchmark matrix and selection details.

\paragraph{Once in comments, register matters.}
Prefilling an empty \code{\#} hurts, an algorithm description helps, and \code{\# Verified approach:} causes register lock-in (Appendix~\ref{app:after}).
Similarly, LiveCodeBench's stdin template plants \code{\# YOUR CODE HERE}, which small models echo (Appendix~\ref{app:lcb}).

\section{Conclusion}
\label{sec:conclusion}
Code LMs write \ph{} when the document around a function looks like an exercise: they copy neighbouring stubs and read file-level text as a genre cue, at the rate real files do.
In an attention knock-out, the header's effect depends on first-half attention and the neighbour's on second-half attention, and both reach a late comment-or-code decision.
Using this structure, targeted representation-level interventions outperform coarse comment bans in two instruction-tuned settings where useful comments make global suppression brittle.

\clearpage
\section*{Limitations}
We evaluate Python function completion with English text; MultiPL-E results are first lines only and not executed.
The behaviour is specific to completion-style prompts: under a chat template with free answers it disappears, and we test three instruct models in that setting.
Fill-in-the-middle is tested on first-line deferral only, on four models and three suffix functions, without pass@1; we do not test repository-level completion, and the real-file experiment is our closest proxy.
Per-model HumanEval tests have limited power ($n{=}164$); the pooled mixed model depends on GrammarCoder once both \model{Coder-1.5B} checkpoints are removed (Section~\ref{sec:default}).
The deferral classifier is a regular expression. On a stratified sheet of 213 distinct first lines, labelled blind by one human annotator and two LLMs (Grok 4.7 and GPT-5.6; Krippendorff's $\alpha{=}0.97$), it matches the majority label on 209 lines (98.1\%) and finds 77 of 80 deferrals; its four errors are all on the placeholder--contentful boundary. The sheet samples distinct strings, so this is not an error rate over completions.
Neighbours and headers come from a small hand-written set; the minimal pairs, MBPP+, and real files reduce but do not remove this dependence.
Some steering configurations were selected from sweeps, so intervention gains are specific to the reported model--setting pairs rather than evidence of uniform superiority; the attention-knockout split by network half is exploratory, and the FIM head comparison is correlational.
Full-generation experiments use models of at most 14B parameters, and mechanistic analyses cover three models of at most \model{Coder-7B}.
The training corpora of most models are not public; our corpus reference is a 10,000-file deduplicated GitHub sample, which under-represents near-duplicate course templates, so the origin of the default tendency remains open.
A benchmark prompt audit (Appendix~\ref{app:lcb}) shows that one benchmark's prompt plants the placeholder itself, so its ban gain is not evidence for the effect studied here.

\section*{Ethics Statement}
This work analyses open-weight code LMs and edits them only at decode time for diagnosis and targeted mitigation.
It introduces no new capability.

\bibliography{paper}

@misc{chen2021humaneval,
  title         = {Evaluating Large Language Models Trained on Code},
  author        = {Chen, Mark and Tworek, Jerry and Jun, Heewoo and Yuan, Qiming and Pinto, Henrique Ponde de Oliveira and Kaplan, Jared and Edwards, Harri and Burda, Yuri and Joseph, Nicholas and Brockman, Greg and Ray, Alex and Puri, Raul and Krueger, Gretchen and Petrov, Michael and Khlaaf, Heidy and Sastry, Girish and Mishkin, Pamela and Chan, Brooke and Gray, Scott and Ryder, Nick and Pavlov, Mikhail and Power, Alethea and Kaiser, Lukasz and Bavarian, Mohammad and Winter, Clemens and Tillet, Philippe and Such, Felipe Petroski and Cummings, Dave and Plappert, Matthias and Chantzis, Fotios and Barnes, Elizabeth and Herbert-Voss, Ariel and Guss, William Hebgen and Nichol, Alex and Paino, Alex and Tezak, Nikolas and Tang, Jie and Babuschkin, Igor and Balaji, Suchir and Jain, Shantanu and Saunders, William and Hesse, Christopher and Carr, Andrew N. and Leike, Jan and Achiam, Josh and Misra, Vedant and Morikawa, Evan and Radford, Alec and Knight, Matthew and Brundage, Miles and Murati, Mira and Mayer, Katie and Welinder, Peter and McGrew, Bob and Amodei, Dario and McCandlish, Sam and Sutskever, Ilya and Zaremba, Wojciech},
  year          = {2021},
  eprint        = {2107.03374},
  archivePrefix = {arXiv},
  primaryClass  = {cs.LG},
  url           = {https://arxiv.org/abs/2107.03374}
}

@misc{alain2016understanding,
  title         = {Understanding intermediate layers using linear classifier probes},
  author        = {Alain, Guillaume and Bengio, Yoshua},
  year          = {2016},
  eprint        = {1610.01644},
  archivePrefix = {arXiv},
  primaryClass  = {stat.ML},
  url           = {https://arxiv.org/abs/1610.01644}
}

@misc{turner2023activation,
  title         = {Steering Language Models With Activation Engineering},
  author        = {Turner, Alexander Matt and Thiergart, Lisa and Leech, Gavin and Udell, David and Vazquez, Juan J. and Mini, Ulisse and MacDiarmid, Monte},
  year          = {2024},
  eprint        = {2308.10248},
  archivePrefix = {arXiv},
  primaryClass  = {cs.CL},
  url           = {https://arxiv.org/abs/2308.10248}
}

@inproceedings{rimsky2024steering,
  title     = {Steering Llama 2 via Contrastive Activation Addition},
  author    = {Rimsky, Nina and Gabrieli, Nick and Schulz, Julian and Tong, Meg and Hubinger, Evan and Turner, Alexander},
  booktitle = {Proceedings of the 62nd Annual Meeting of the Association for Computational Linguistics (Volume 1: Long Papers)},
  month     = aug,
  year      = {2024},
  address   = {Bangkok, Thailand},
  publisher = {Association for Computational Linguistics},
  url       = {https://aclanthology.org/2024.acl-long.828/},
  doi       = {10.18653/v1/2024.acl-long.828},
  pages     = {15504--15522}
}

@misc{zou2023representation,
  title         = {Representation Engineering: A Top-Down Approach to {AI} Transparency},
  author        = {Zou, Andy and Phan, Long and Chen, Sarah and Campbell, James and Guo, Phillip and Ren, Richard and Pan, Alexander and Yin, Xuwang and Mazeika, Mantas and Dombrowski, Ann-Kathrin and Goel, Shashwat and Li, Nathaniel and Byun, Michael J. and Wang, Zifan and Mallen, Alex and Basart, Steven and Koyejo, Sanmi and Song, Dawn and Fredrikson, Matt and Kolter, J. Zico and Hendrycks, Dan},
  year          = {2025},
  eprint        = {2310.01405},
  archivePrefix = {arXiv},
  primaryClass  = {cs.LG},
  url           = {https://arxiv.org/abs/2310.01405}
}

@inproceedings{geng2023grammar,
  title     = {Grammar-Constrained Decoding for Structured {NLP} Tasks without Finetuning},
  author    = {Geng, Saibo and Josifoski, Martin and Peyrard, Maxime and West, Robert},
  booktitle = {Proceedings of the 2023 Conference on Empirical Methods in Natural Language Processing},
  month     = dec,
  year      = {2023},
  address   = {Singapore},
  publisher = {Association for Computational Linguistics},
  url       = {https://aclanthology.org/2023.emnlp-main.674/},
  doi       = {10.18653/v1/2023.emnlp-main.674},
  pages     = {10932--10952}
}

@inproceedings{scholak2021picard,
  title     = {{PICARD}: Parsing Incrementally for Constrained Auto-Regressive Decoding from Language Models},
  author    = {Scholak, Torsten and Schucher, Nathan and Bahdanau, Dzmitry},
  booktitle = {Proceedings of the 2021 Conference on Empirical Methods in Natural Language Processing},
  month     = nov,
  year      = {2021},
  address   = {Online and Punta Cana, Dominican Republic},
  publisher = {Association for Computational Linguistics},
  url       = {https://aclanthology.org/2021.emnlp-main.779/},
  doi       = {10.18653/v1/2021.emnlp-main.779},
  pages     = {9895--9901}
}

@misc{nostalgebraist2020logitlens,
  title        = {Interpreting {GPT}: The Logit Lens},
  author       = {{nostalgebraist}},
  year         = {2020},
  howpublished = {\url{https://www.lesswrong.com/posts/AcKRB8wDpdaN6v6ru/interpreting-gpt-the-logit-lens}},
  note         = {LessWrong}
}

@inproceedings{chen2024mango,
  title     = {Comments as Natural Logic Pivots: Improve Code Generation via Comment Perspective},
  author    = {Chen, Yijie and Liu, Yijin and Meng, Fandong and Chen, Yufeng and Xu, Jinan and Zhou, Jie},
  booktitle = {Findings of the Association for Computational Linguistics: ACL 2024},
  month     = aug,
  year      = {2024},
  address   = {Bangkok, Thailand},
  publisher = {Association for Computational Linguistics},
  url       = {https://aclanthology.org/2024.findings-acl.420/},
  doi       = {10.18653/v1/2024.findings-acl.420},
  pages     = {7040--7051}
}

@article{jiang2024selfplanning,
  title   = {Self-Planning Code Generation with Large Language Models},
  author  = {Jiang, Xue and Dong, Yihong and Wang, Lecheng and Zheng, Fang and Shang, Qiwei and Li, Ge and Jin, Zhi and Jiao, Wenpin},
  journal = {ACM Transactions on Software Engineering and Methodology},
  volume  = {33},
  number  = {7},
  pages   = {182:1--182:30},
  year    = {2024},
  doi     = {10.1145/3672456},
  publisher = {ACM}
}

@article{li2023scot,
  title   = {Structured Chain-of-Thought Prompting for Code Generation},
  author  = {Li, Jia and Li, Ge and Li, Yongmin and Jin, Zhi},
  journal = {ACM Transactions on Software Engineering and Methodology},
  volume  = {34},
  number  = {2},
  pages   = {37:1--37:23},
  year    = {2025},
  doi     = {10.1145/3690635},
  publisher = {ACM}
}

@misc{ma2023secot,
  title         = {Bridging Code Semantic and {LLMs}: Semantic Chain-of-Thought Prompting for Code Generation},
  author        = {Ma, Yingwei and Yu, Yue and Li, Shanshan and Jiang, Yu and Guo, Yong and Zhang, Yuanliang and Xie, Yutao and Liao, Xiangke},
  year          = {2023},
  eprint        = {2310.10698},
  archivePrefix = {arXiv},
  primaryClass  = {cs.CL},
  url           = {https://arxiv.org/abs/2310.10698}
}

@inproceedings{song2024codeneedscomments,
  title     = {Code Needs Comments: Enhancing Code {LLM}s with Comment Augmentation},
  author    = {Song, Demin and Guo, Honglin and Zhou, Yunhua and Xing, Shuhao and Wang, Yudong and Song, Zifan and Zhang, Wenwei and Guo, Qipeng and Yan, Hang and Qiu, Xipeng and Lin, Dahua},
  booktitle = {Findings of the Association for Computational Linguistics: ACL 2024},
  month     = aug,
  year      = {2024},
  address   = {Bangkok, Thailand},
  publisher = {Association for Computational Linguistics},
  url       = {https://aclanthology.org/2024.findings-acl.809/},
  doi       = {10.18653/v1/2024.findings-acl.809},
  pages     = {13640--13656}
}

@inproceedings{lei2025planning,
  title     = {Planning-Driven Programming: A Large Language Model Programming Workflow},
  author    = {Lei, Chao and Chang, Yanchuan and Lipovetzky, Nir and Ehinger, Krista A.},
  booktitle = {Proceedings of the 63rd Annual Meeting of the Association for Computational Linguistics (Volume 1: Long Papers)},
  month     = jul,
  year      = {2025},
  address   = {Vienna, Austria},
  publisher = {Association for Computational Linguistics},
  url       = {https://aclanthology.org/2025.acl-long.621/},
  doi       = {10.18653/v1/2025.acl-long.621},
  pages     = {12647--12684}
}

@inproceedings{ghoshal2026thinkless,
  title     = {Think Less, Code Better: Probing When Chain-of-Thought Hurts and How to Route Around It},
  author    = {Ghoshal, Rajarshi and Basak, Debadri and Abdelhalim, Salma E. and Arora, Pratibha K.},
  booktitle = {Proceedings of the 64th Annual Meeting of the Association for Computational Linguistics (Volume 4: Student Research Workshop)},
  month     = jul,
  year      = {2026},
  address   = {San Diego, California, United States},
  publisher = {Association for Computational Linguistics},
  url       = {https://aclanthology.org/2026.acl-srw.13/},
  doi       = {10.18653/v1/2026.acl-srw.13},
  pages     = {152--162}
}

@misc{pan2025talking,
  title         = {Talking to Itself While Coding: What Makes Comments Help Code Generation?},
  author        = {Pan, Dangfeng and Sun, Zhensu and Zhang, Cenyuan and Lo, David and Du, Xiaoning},
  year          = {2026},
  eprint        = {2609.09242},
  archivePrefix = {arXiv},
  primaryClass  = {cs.SE},
  url           = {https://arxiv.org/abs/2609.09242},
  note          = {Accepted to Findings of EMNLP 2026}
}

@inproceedings{macke2024documentation,
  title     = {Testing the Effect of Code Documentation on Large Language Model Code Understanding},
  author    = {Macke, William and Doyle, Michael},
  booktitle = {Findings of the Association for Computational Linguistics: NAACL 2024},
  month     = jun,
  year      = {2024},
  address   = {Mexico City, Mexico},
  publisher = {Association for Computational Linguistics},
  url       = {https://aclanthology.org/2024.findings-naacl.66/},
  doi       = {10.18653/v1/2024.findings-naacl.66},
  pages     = {1044--1050}
}

@inproceedings{obrien2024todo,
  title     = {Are Prompt Engineering and {TODO} Comments Friends or Foes? An Evaluation on {GitHub} {Copilot}},
  author    = {O'Brien, David and Biswas, Sumon and Imtiaz, Sayem Mohammad and Abdalkareem, Rabe and Shihab, Emad and Rajan, Hridesh},
  booktitle = {Proceedings of the IEEE/ACM 46th International Conference on Software Engineering},
  year      = {2024},
  pages     = {219:1--219:13},
  publisher = {ACM},
  doi       = {10.1145/3597503.3639176}
}

@inproceedings{dinh2023buggy,
  title     = {Large Language Models of Code Fail at Completing Code with Potential Bugs},
  author    = {Dinh, Tuan and Zhao, Jinman and Tan, Samson and Negrinho, Renato and Lausen, Leonard and Zha, Sheng and Karypis, George},
  booktitle = {Advances in Neural Information Processing Systems},
  volume    = {36},
  year      = {2023}
}

@inproceedings{wang2023recode,
  title     = {{ReCode}: Robustness Evaluation of Code Generation Models},
  author    = {Wang, Shiqi and Li, Zheng and Qian, Haifeng and Yang, Chenghao and Wang, Zijian and Shang, Mingyue and Kumar, Varun and Tan, Samson and Ray, Baishakhi and Bhatia, Parminder and Nallapati, Ramesh and Ramanathan, Murali Krishna and Roth, Dan and Xiang, Bing},
  booktitle = {Proceedings of the 61st Annual Meeting of the Association for Computational Linguistics (Volume 1: Long Papers)},
  year      = {2023},
  publisher = {Association for Computational Linguistics},
  url       = {https://aclanthology.org/2023.acl-long.773/},
  pages     = {13818--13843}
}

@inproceedings{sclar2024quantifying,
  title     = {Quantifying Language Models' Sensitivity to Spurious Features in Prompt Design or: How {I} Learned to Start Worrying about Prompt Formatting},
  author    = {Sclar, Melanie and Choi, Yejin and Tsvetkov, Yulia and Suhr, Alane},
  booktitle = {The Twelfth International Conference on Learning Representations},
  year      = {2024}
}

@inproceedings{liu2024repobench,
  title     = {{RepoBench}: Benchmarking Repository-Level Code Auto-Completion Systems},
  author    = {Liu, Tianyang and Xu, Canwen and McAuley, Julian},
  booktitle = {The Twelfth International Conference on Learning Representations},
  year      = {2024}
}

@inproceedings{ding2023crosscodeeval,
  title     = {{CrossCodeEval}: A Diverse and Multilingual Benchmark for Cross-File Code Completion},
  author    = {Ding, Yangruibo and Wang, Zijian and Ahmad, Wasi Uddin and Ding, Hantian and Tan, Ming and Jain, Nihal and Ramanathan, Murali Krishna and Nallapati, Ramesh and Bhatia, Parminder and Roth, Dan and Xiang, Bing},
  booktitle = {Advances in Neural Information Processing Systems (Datasets and Benchmarks Track)},
  volume    = {36},
  year      = {2023}
}

@article{huang2026commenttraps,
  title     = {Comment Traps: How Defective Commented-Out Code Augment Defects in {AI}-Assisted Code Generation},
  author    = {Huang, Yuan and Zhou, Yukang and Chen, Xiangping and Zheng, Zibin},
  journal   = {Proceedings of the ACM on Software Engineering},
  volume    = {3},
  number    = {FSE},
  pages     = {1650--1670},
  year      = {2026},
  doi       = {10.1145/3797102},
  publisher = {ACM},
  eprint    = {2512.20334},
  archivePrefix = {arXiv},
  primaryClass  = {cs.SE}
}

@misc{olsson2022induction,
  title         = {In-context Learning and Induction Heads},
  author        = {Olsson, Catherine and Elhage, Nelson and Nanda, Neel and Joseph, Nicholas and DasSarma, Nova and Henighan, Tom and Mann, Ben and Askell, Amanda and Bai, Yuntao and Chen, Anna and Conerly, Tom and Drain, Dawn and Ganguli, Deep and Hatfield-Dodds, Zac and Hernandez, Danny and Johnston, Scott and Jones, Andy and Kernion, Jackson and Lovitt, Liane and Ndousse, Kamal and Amodei, Dario and Brown, Tom and Clark, Jack and Kaplan, Jared and McCandlish, Sam and Olah, Chris},
  year          = {2022},
  eprint        = {2209.11895},
  archivePrefix = {arXiv},
  primaryClass  = {cs.LG},
  url           = {https://arxiv.org/abs/2209.11895}
}

@inproceedings{meng2022locating,
  title     = {Locating and Editing Factual Associations in {GPT}},
  author    = {Meng, Kevin and Bau, David and Andonian, Alex and Belinkov, Yonatan},
  booktitle = {Advances in Neural Information Processing Systems},
  volume    = {35},
  year      = {2022}
}

@inproceedings{wang2023ioi,
  title     = {Interpretability in the Wild: a Circuit for Indirect Object Identification in {GPT}-2 Small},
  author    = {Wang, Kevin Ro and Variengien, Alexandre and Conmy, Arthur and Shlegeris, Buck and Steinhardt, Jacob},
  booktitle = {The Eleventh International Conference on Learning Representations},
  year      = {2023}
}

@inproceedings{zhang2024patching,
  title     = {Towards Best Practices of Activation Patching in Language Models: Metrics and Methods},
  author    = {Zhang, Fred and Nanda, Neel},
  booktitle = {The Twelfth International Conference on Learning Representations},
  year      = {2024}
}

@book{biber1988variation,
  title     = {Variation across Speech and Writing},
  author    = {Biber, Douglas},
  publisher = {Cambridge University Press},
  address   = {Cambridge},
  year      = {1988},
  doi       = {10.1017/CBO9780511621024}
}

@misc{kocetkov2022stack,
  title         = {The Stack: 3 {TB} of Permissively Licensed Source Code},
  author        = {Kocetkov, Denis and Li, Raymond and Ben Allal, Loubna and Li, Jia and Mou, Chenghao and Mu{\~n}oz Ferrandis, Carlos and Jernite, Yacine and Mitchell, Margaret and Hughes, Sean and Wolf, Thomas and Bahdanau, Dzmitry and von Werra, Leandro and de Vries, Harm},
  year          = {2022},
  eprint        = {2211.15533},
  archivePrefix = {arXiv},
  primaryClass  = {cs.CL},
  url           = {https://arxiv.org/abs/2211.15533}
}

@article{cassano2023multipl,
  title   = {{MultiPL-E}: A Scalable and Polyglot Approach to Benchmarking Neural Code Generation},
  author  = {Cassano, Federico and Gouwar, John and Nguyen, Daniel and Nguyen, Sydney and Phipps-Costin, Luna and Pinckney, Donald and Yee, Ming-Ho and Zi, Yangtian and Anderson, Carolyn Jane and Feldman, Molly Q. and Guha, Arjun and Greenberg, Michael and Jangda, Abhinav},
  journal = {IEEE Transactions on Software Engineering},
  volume  = {49},
  number  = {7},
  pages   = {3675--3691},
  year    = {2023},
  doi     = {10.1109/TSE.2023.3267446}
}

@inproceedings{liu2023evalplus,
  title     = {Is Your Code Generated by {ChatGPT} Really Correct? Rigorous Evaluation of Large Language Models for Code Generation},
  author    = {Liu, Jiawei and Xia, Chunqiu Steven and Wang, Yuyang and Zhang, Lingming},
  booktitle = {Advances in Neural Information Processing Systems},
  volume    = {36},
  year      = {2023}
}

\appendix

\section{Comment content after the opener}
\label{app:after}
We prefill the first body line and let the model continue (Table~\ref{tab:after}).
An empty \code{\#} and a generic plan hurt both the \model{Coder-1.5B} and the \model{Coder-7B} coder ($-15.9$ to $-32.3$\pp{}$^*$), and so does a MANGO-style instruction to write logical comments first (\model{Coder-7B} $-11.0$\pp{}$^*$).
One English sentence that describes the reference algorithm helps \model{Coder-1.5B} ($+9.1$\pp{}$^*$) and \model{DS-1.3B} ($+9.1$\pp{}$^*$).
Writing the canonical first line \emph{as code} helps all three models ($+25.6$, $+14.0$, $+17.7$\pp{}$^*$).
The same line written as \code{\# Verified approach: <line>} hurts all three ($-12.8$, $-23.2$, $-10.4$\pp{}$^*$); $81$, $73$, and $118$ of $164$ completions never write a code line and repeat comments instead.
A bare \code{\# <line>} is in between ($-3.7$ to $-4.9$\pp{}, n.s.), so the label, not the comment marker alone, keeps the model in comment mode.

\begin{table}[t]
\centering
\small
\setlength{\tabcolsep}{3pt}
\resizebox{\columnwidth}{!}{%
\begin{tabular}{lrrr}
\toprule
Prefilled first line & \model{Coder-1.5B} & \model{Coder-7B} & DS-1.3B \\
\midrule
allow (pass@1) & 42.1 & 59.8 & 29.3 \\
\midrule
empty \code{\#} & $-15.9^*$ & $-18.3^*$ & -- \\
generic plan comment & $-26.2^*$ & $-32.3^*$ & -- \\
``write logical comments'' & $-7.3$ & $-11.0^*$ & -- \\
algorithm in one sentence & $+9.1^*$ & $+6.7$ & $+9.1^*$ \\
\midrule
canonical line as code & $+25.6^*$ & $+14.0^*$ & $+17.7^*$ \\
\code{\# <line>} & $-3.7$ & $-3.7$ & $-4.9$ \\
\code{\# Verified approach: <line>} & $-12.8^*$ & $-23.2^*$ & $-10.4^*$ \\
\bottomrule
\end{tabular}%
}
\caption{\textbf{Content and register decide the sign.} $\Delta$\pp{} vs.\ allow on HumanEval. This harness has \model{Coder-1.5B} allow $42.1\%$ (HF \code{generate}); compare only within a column.}
\label{tab:after}
\end{table}

\section{Models, default rates, and header results}
\label{app:models}
Table~\ref{tab:p0} lists the 24 checkpoints under the plain HumanEval prompt.
Placeholder: greedy first body line is a placeholder comment.
Allow and ban\_ph: HumanEval pass@1 without and with the placeholder-only ban; R/B: rescues/breaks; $p$: exact McNemar.
Java/C++/JS: placeholder rate of the first body line on MultiPL-E ($n{=}158/161/161$), not executed.
$^\dagger$: skips the body at a token boundary under the HumanEval prompt.
Table~\ref{tab:p2full} gives the header result for every model.

\begin{table*}[t]
\centering
\small
\setlength{\tabcolsep}{4pt}
\begin{tabular}{llrrrrrrrr}
\toprule
Model & Family & Placeholder & Allow & ban\_ph $\Delta$ & R/B & $p$ & Java & C++ & JS \\
\midrule
Qwen2.5-Coder-0.5B & Qwen2.5-Coder & 29.9 & 28.7 & $-0.6$ & 2/3 & 1.00 & 22.2 & 1.2 & 0.6 \\
Qwen2.5-Coder-1.5B & Qwen2.5-Coder & 34.8 & 42.7 & $+7.3^*$ & 14/2 & 0.004 & 6.3 & 4.3 & 0.6 \\
Qwen2.5-\model{Coder-1.5B-IT} & Qwen2.5-Coder & 14.0 & 59.1 & $+5.5^*$ & 9/0 & 0.004 & 3.8 & 0.0 & 0.0 \\
Qwen2.5-Coder-3B & Qwen2.5-Coder & 10.4 & 50.0 & $+0.6$ & 4/3 & 1.00 & 29.1 & 3.1 & 0.0 \\
Qwen2.5-Coder-7B & Qwen2.5-Coder & 9.1 & 61.0 & $+1.8$ & 4/1 & 0.38 & 16.5 & 1.9 & 0.0 \\
Qwen2.5-7B & Qwen2.5 & 19.5 & 56.1 & $+2.4$ & 6/2 & 0.29 & 13.3 & 1.2 & 0.6 \\
Qwen3-4B & Qwen3 & 12.2 & 72.0 & $+1.8$ & 4/1 & 0.38 & 34.2 & 19.3 & 24.2 \\
Qwen3-8B & Qwen3 & 4.3 & 67.7 & $0.0$ & 0/0 & 1.00 & 3.8 & 1.2 & 0.6 \\
Qwen3-8B-Instruct & Qwen3 & 7.9 & 71.3 & $-1.2$ & 0/2 & 0.50 & 41.1 & 17.4 & 28.6 \\
Qwen3-14B & Qwen3 & 5.5 & 72.6 & $-0.6$ & 1/2 & 1.00 & 6.3 & 3.7 & 0.0 \\
Qwen3.5-0.8B & Qwen3.5 & 3.0 & 22.0 & $+0.6$ & 1/0 & 1.00 & 11.4 & 0.0 & 1.2 \\
Qwen3.5-4B & Qwen3.5 & 6.7 & 51.2 & $0.0$ & 0/0 & 1.00 & 7.0 & 0.0 & 0.0 \\
DeepSeek-Coder-1.3B & DeepSeek-Coder & 24.4 & 29.3 & $+1.8$ & 3/0 & 0.25 & 31.6 & 6.8 & 0.6 \\
Seed-Coder-8B & Seed-Coder & 0.0 & 81.1 & $0.0$ & 0/0 & 1.00 & 0.0 & 0.0 & 0.0 \\
CodeGemma-2B & CodeGemma & 23.2 & 32.9 & $+2.4$ & 5/1 & 0.22 & 66.5 & 34.8 & 16.1 \\
Gemma3-4B & Gemma & 11.6 & 35.4 & $+0.6$ & 1/0 & 1.00 & 7.0 & 0.6 & 0.0 \\
Gemma4-E2B & Gemma & 20.7 & 26.2 & $+1.2$ & 3/1 & 0.63 & 31.0 & 2.5 & 14.3 \\
Gemma4-4B & Gemma & 38.4 & 34.1 & $+1.8$ & 7/4 & 0.55 & 34.2 & 6.2 & 6.8 \\
Llama-3.1-8B & Llama-3.1 & 3.7 & 33.5 & $+1.2$ & 2/0 & 0.50 & 24.7 & 1.2 & 0.6 \\
OpenCoder-1.5B & OpenCoder & 6.1 & 52.4 & $+0.6$ & 1/0 & 1.00 & 0.0 & 0.6 & 0.6 \\
GrammarCoder-1.5B & GrammarCoder & 79.9 & 5.5 & $+21.3^*$ & 35/0 & $6{\times}10^{-11}$ & 86.7 & 55.3 & 43.5 \\
StarCoder2-3B$^\dagger$ & StarCoder2 & 0.0 & 13.4 & $0.0$ & 0/0 & 1.00 & 15.8 & 1.2 & 1.9 \\
StarCoder2-7B$^\dagger$ & StarCoder2 & 0.0 & 9.1 & $0.0$ & 0/0 & 1.00 & 31.6 & 3.1 & 0.0 \\
StableCode-3B$^\dagger$ & StableCode & 0.0 & 1.2 & $0.0$ & 0/0 & 1.00 & 3.8 & 0.0 & 1.2 \\
\bottomrule
\end{tabular}
\caption{\textbf{Plain HumanEval prompt, 24 checkpoints.} Rates and pass@1 in \%, $\Delta$ in \pp{}.}
\label{tab:p0}
\end{table*}

\begin{table*}[t]
\centering
\small
\setlength{\tabcolsep}{3.8pt}
\begin{tabular}{lrrrrrr|lrrrrrr}
\toprule
Model & none & neut. & ex. & ref. & $\Delta$ & 5/5 & Model & none & neut. & ex. & ref. & $\Delta$ & 5/5 \\
\midrule
Coder-0.5B & 30.5 & 26.2 & 55.6 & 35.1 & $+20.5$ & \checkmark & Qwen3-14B & 5.5 & 8.5 & 61.8 & 2.0 & $+59.9$ & \checkmark \\
Coder-1.5B & 36.0 & 16.8 & 51.6 & 27.4 & $+24.1$ & \checkmark & Qwen3-30B-A3B & 6.7 & 4.3 & 75.5 & 3.8 & $+71.7$ & \checkmark \\
\model{Coder-1.5B-IT} & 14.0 & 11.0 & 44.3 & 13.2 & $+31.1$ & \checkmark & Qwen3.5-0.8B & 3.0 & 4.0 & 24.0 & 8.3 & $+15.7$ & \checkmark \\
Coder-3B & 11.6 & 11.9 & 62.0 & 17.2 & $+44.8$ & 4/5 & Qwen3.5-4B & 6.7 & 12.2 & 68.7 & 18.2 & $+50.5$ & \checkmark \\
Coder-7B & 11.6 & 20.7 & 59.0 & 15.6 & $+43.4$ & \checkmark & DeepSeek-1.3B & 28.7 & 24.7 & 62.0 & 40.9 & $+21.1$ & \checkmark \\
Qwen2.5-7B & 20.7 & 20.1 & 62.0 & 23.4 & $+38.5$ & \checkmark & Seed-Coder-8B & 0.0 & 6.4 & 35.4 & 3.5 & $+31.8$ & \checkmark \\
Qwen3-4B & 12.2 & 17.4 & 61.3 & 6.5 & $+54.9$ & \checkmark & CodeGemma-2B & 26.8 & 23.5 & 76.2 & 27.3 & $+48.9$ & \checkmark \\
Qwen3-8B & 4.9 & 8.5 & 49.6 & 7.0 & $+42.7$ & \checkmark & Gemma3-4B & 12.2 & 14.6 & 52.1 & 27.4 & $+24.6$ & \checkmark \\
Qwen3-8B-IT & 8.5 & 7.0 & 62.4 & 2.9 & $+59.5$ & \checkmark & Gemma4-E2B & 20.7 & 20.7 & 79.4 & 41.8 & $+37.6$ & \checkmark \\
Llama-3.1-8B & 3.7 & 4.3 & 19.9 & 6.0 & $+13.9$ & \checkmark & Gemma4-4B & 38.4 & 36.3 & 89.4 & 38.8 & $+50.6$ & \checkmark \\
CodeLlama-7B-Py & 20.1 & 28.4 & 72.6 & 43.8 & $+28.8$ & \checkmark & OpenCoder-1.5B & 12.2 & 23.5 & 61.2 & 33.0 & $+28.2$ & \checkmark \\
GrammarCoder-1.5B & 83.5 & 84.1 & 89.8 & 74.6 & $+15.1$ & \checkmark & & & & & & & \\
\bottomrule
\end{tabular}
\caption{\textbf{File headers, all 23 comparable models.} First-line deferral (\%); every $\Delta$ is significant (task-level permutation test, Holm-corrected; task-bootstrap 95\% CIs exclude 0 for all 23). 5/5: exercise above reference on all five paraphrase pairs. The plain-prompt rate (none) here is from the same first-line run and can differ slightly from Table~\ref{tab:p0}.}
\label{tab:p2full}
\end{table*}

\section{Prompt strings}
\label{app:prompts}
\paragraph{Neighbours.}
Three functions, each with a signature and a doctest docstring: \code{count\_vowels(text)} (finished body \code{return sum(1 for ch in text.lower() if ch in 'aeiou')}), \code{merge\_intervals(intervals)} (an eight-line loop), and \code{is\_palindrome(text)} (\code{return text == text[::-1]}).
The stub bodies replace the finished body with \ph{} + \code{pass}, \code{\# TODO: implement} + \code{pass}, \code{raise NotImplementedError}, \code{...}, or \code{pass}; the no-body variant ends after the docstring.
The neighbour is followed by a blank line and the HumanEval prompt.

\paragraph{File headers.}
Exercise: \code{\# homework/hw3\_starter.py --- fill in the functions}; \code{\# Exercise 4: implement each function below so that the doctests pass.}; \code{\# lab\_05/starter\_code.py} + \code{\# Students: complete the functions below.}; \code{"""Assignment 2 skeleton. Do not change the function signatures."""}; \code{\# templates/problem\_set.py (starter template)}.
Reference: \code{\# solutions/reference.py}; \code{\# Reference implementations. All functions are complete and pass their tests.}; \code{\# src/core/helpers.py} + \code{\# Production module; every function is implemented and unit-tested.}; \code{"""Worked solutions for the problem set, with full implementations."""}; \code{\# answer\_key.py --- instructor solutions}.
Neutral: \code{\# utils.py}; \code{\# -*- coding: utf-8 -*-}.
Pair $j$ is exercise $j$ with reference $j$.

\begin{figure}[t]
\centering
\includegraphics[width=\columnwidth]{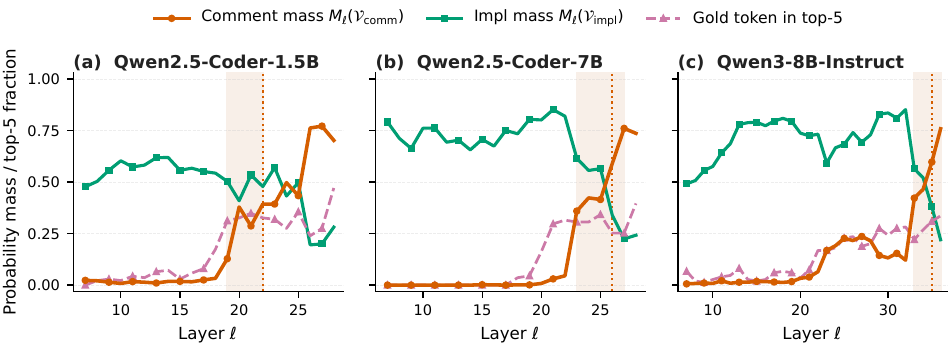}
\caption{\textbf{Logit lens at the body start} (default prompt, triggered tasks).
Comment-opener mass (orange) stays low through mid layers and surges in the last third, in parallel with implementation mass (green).}
\label{fig:lens}
\end{figure}

\section{Error taxonomy}
\label{app:errors}
On \model{Coder-1.5B} (HF harness, allow $42.1\%$), 65 of 95 allow failures write a comment and then wrong code; post-hoc deletion of TODO comments and a matched random token mask leave pass@1 at allow; mean completion length rises from 19.9 to 24.7 lines under the ban.

\section{Representation-Level Intervention Matrix}
\label{app:steer}
Table~\ref{tab:gated_editor} gives the full matrix behind the targeted comparisons in Section~\ref{sec:discussion}.
On HumanEval, subtracting the separating direction $d$ at \model{Coder-1.5B} layer 22 achieves $+6.1$\pp{} ($20/10$, $p{=}0.10$), and a leave-one-task-out harm direction yields $+6.1$\pp{}$^*$ ($14/4$).
On complex multi-line code (BigCodeBench-hard complete, $n{=}148$, scored with the official evaluator), coarse \code{ban} changes pass@1 by $-2.0$\pp{} on \model{Coder-1.5B}, whereas the linear probe-gated residual editor ($\tau{=}0.5$, CV AUC $\approx 0.99$) yields $+1.4$\pp{}; neither change nor their direct difference is significant.
On \model{Coder-7B}, whitespace-gated editing yields $+4.7$\pp{} against coarse \code{ban}'s $+2.0$\pp{}, again without a significant direct difference.

On \model{Qwen3-8B-Instruct}, coarse \code{ban} breaks 13 tasks that contain valid comment reasoning ($13/13$ rescues/breaks, net $0.0$\pp{}), whereas latent steering at layer 35 gains $+5.5$\pp{} ($16/7$, $p{=}0.093$), protecting 6 tasks broken by coarse \code{ban} and uniquely rescuing 4 tasks without touching surface tokens. Conversely, adding $d$ yields explanatory comments rather than placeholders (29 contentful vs.\ 4 placeholder flips at $\alpha{=}0.25$) with no change in pass@1.

\paragraph{Document gate and inference efficiency.}
To avoid representation drift on natural code, we train a linear document classifier at prompt-end hidden states (gate AUC $0.898$ on \model{Coder-1.5B}, $0.995$ on \model{DS-1.3B}, $0.999$ on \model{Coder-7B}). The gate fires on $\le 1\%$ of clean prompts. In difficult skeleton environments, document-gated editing adds significant accuracy over pure token constraints (\model{DS-1.3B} under \code{...}: $+7.9$\pp{}$^*$, 19/6; \model{Coder-7B} under exercise header: $+10.4$\pp{}$^*$, 24/7).
In our A100 profile, evaluating linear probe dot products ($1\times D$) at statement starts achieves $45.6$ tok/s ($1.14\times$ relative to allow's $40.1$ tok/s), whereas updating large vocabulary logits vectors at every step slows coarse \code{ban} to $34.5$ tok/s ($0.86\times$).

\begin{table}[h]
\centering
\small
\setlength{\tabcolsep}{3pt}
\resizebox{\columnwidth}{!}{%
\begin{tabular}{llrrrr}
\toprule
Dataset & Model & Allow & Coarse \code{ban} & WS-Gated & Probe-Gated ($\tau{=}0.5$) \\
\midrule
HumanEval & \model{Coder-1.5B} & 42.1 & $+7.9^*$ & $+6.1$ & $+5.5$ \\
 & \model{Coder-1.5B-IT} & 45.7 & $+11.0^*$ & $\mathbf{+19.5^*}$ & $\mathbf{+17.7^*}$ \\
 & \model{Coder-7B} & 59.8 & $+1.2$ & $+2.4$ & $+1.8$ \\
 & \model{DS-1.3B} & 29.3 & $+2.4$ & $+3.0$ & $+2.4$ \\
 & \model{Qwen3-8B-IT} & 68.3 & $0.0$ & $+5.5$ & $+4.3$ \\
 & \model{Seed-Coder-8B} & 78.7 & $-2.4$ & $-2.4$ & $-1.8$ \\
 & \model{Gemma4-E2B} & 26.2 & $+1.2$ & $-0.6$ & $+1.8$ \\
\midrule
BigCodeBench & \model{Coder-1.5B} & 12.8 & $-2.0$ & $-4.7$ & $\mathbf{+1.4}$ \\
(hard) & \model{Coder-1.5B-IT} & 6.1 & $0.0$ & $\mathbf{+2.0}$ & $\mathbf{+1.3}$ \\
 & \model{Coder-7B} & 17.6 & $+2.0$ & $\mathbf{+4.7}$ & $+0.7$ \\
 & \model{DS-1.3B} & 3.4 & $+0.7$ & $+0.7$ & $\mathbf{+2.0}$ \\
 & \model{Qwen3-8B-IT} & 22.3 & $-2.7$ & $-1.4$ & $-2.0$ \\
 & \model{Seed-Coder-8B} & 33.1 & $0.0$ & $-5.4$ & $0.0$ \\
 & \model{Gemma4-E2B} & 4.7 & $+2.0$ & $-1.4$ & $+1.4$ \\
\bottomrule
\end{tabular}%
}
\caption{\textbf{Representation-level interventions and coarse lexical bans.}
Pass@1 $\Delta$\pp{} relative to allow across seven models on HumanEval ($n{=}164$) and BigCodeBench-hard complete ($n{=}148$).
Stars test each cell against allow; they do not test an intervention directly against \code{ban}. Bold marks the largest point estimate in a row.}
\label{tab:gated_editor}
\end{table}

\section{Head routing}
\label{app:route}
Same prompt pair and log-odds as Figure~\ref{fig:patch}.
For each head, we run the stubbed-neighbour prompt and replace that head's output (the input to the attention output projection) at the decision positions with its value from the finished-neighbour run.
The normalised effect $e$ is the fraction of the neighbour gap in mean log-odds that the patch closes.
The induction score is the attention a head pays to the token after the previous occurrence of the current token on a random sequence repeated twice (50 tokens, 4 sequences); heads scoring $\ge 0.3$ count as induction heads.
Single-head patching understates the circuit: over the 16 readers and 3 writers of Figure~\ref{fig:access}b, the joint patch recovers $1.41$, $1.00$, and $1.53$ times the sum of single-head patches (\model{Coder-1.5B}, \model{DS-1.3B}, \model{Coder-7B}), which is why single-head sums in Table~\ref{tab:route} do not match the layer sums.
Learned one-dimensional interchange directions (DAS) overshoot the full patch for the copy pair; restricted to the span of the observed activation differences, one direction recovers $0.56$--$0.83$ of the header gap and $0.26$--$0.68$ of the copy gap, and we treat the learned values as upper bounds.

\begin{table}[h]
\centering
\small
\setlength{\tabcolsep}{3pt}
\resizebox{\columnwidth}{!}{%
\begin{tabular}{lrrr}
\toprule
 & \model{Coder-1.5B} & DS-1.3B & \model{Coder-7B} \\
\midrule
Neighbour gap in mean log-odds & 7.83 & 7.79 & 9.93 \\
Induction heads (of all) & 15/336 & 13/384 & 20/784 \\
\quad among the top-12 by $e$ & 2 & 1 & 1 \\
\quad share of positive single-head $e$ & 13\% & 7\% & 8\% \\
Spearman($e$, induction score) & $+0.13$ & $+0.04$ & $-0.01$ \\
Largest head & L23.H9 & L9.H1 & L25.H18 \\
\quad its induction score & 0.00 & 0.00 & 0.00 \\
Sum of single-layer attention $e$ & 0.79 & 1.10 & 0.87 \\
Sum of single-head $e$ & 0.12 & 0.70 & $-4.44$ \\
Max $|\cos|$(neighbour direction, $d$) & 0.17 & 0.19 & 0.18 \\
\bottomrule
\end{tabular}%
}
\caption{\textbf{The heads that carry the neighbour are not induction heads.} Neighbour direction: mean difference of last-prompt-position states, stubbed minus finished neighbour.}
\label{tab:route}
\end{table}

\section{Robustness}
\label{app:robust}
Predictions were written before these runs.
\textbf{Sampling} (HumanEval, top-$p$ 0.95, five samples; \model{Coder-1.5B}, \model{Coder-7B}, \model{DS-1.3B}): stubbed minus finished neighbour is $+88.5$ to $+97.3$\pp{} at $T{=}0.2$ and $+84.5$ to $+92.6$\pp{} at $T{=}0.8$; the exercise--reference gap is $+18.8$ to $+44.9$\pp{} and $+11.4$ to $+33.8$\pp{}, with CIs above 0 everywhere.
Other-form deferral stays $\le 0.2\%$ at $T{=}0.2$; at $T{=}0.8$ it reaches $1.3$--$3.2\%$ (\model{DS-1.3B} exceeds the pre-registered $2\%$).
\textbf{MBPP+} ($n{=}378$, HumanEval-style prompts, 25 comparable models): stubbed minus finished $\ge 50$\pp{} on 25/25, largest per-form median other-form deferral $0.0\%$, exercise $>$ reference on 25/25, each significant (median $+30.8$\pp{}).
\textbf{Larger checkpoints}: Qwen2.5-Coder-14B, DeepSeek-Coder-6.7B, and Qwen2.5-Coder-7B-Instruct reproduce copying (median copy $96.5$--$98.6\%$ per form), the exercise--reference gap ($+64.4$, $+22.8$, $+29.5$\pp{}$^*$), and all three minimal-pair cues (9/9 significant).
\textbf{Chat template} (Table~\ref{tab:chat}): with the code prefilled as the start of the answer, the stubbed neighbour still raises deferral but headers barely do; with free answers, deferral is $0\%$ and the models implement the stubbed neighbour as well ($90.7$, $98.3$, $98.8\%$ of answers).
\textbf{Task-clustered tests}: re-testing every header comparison with the task as the unit (bootstrap CI and sign-flip permutation, Holm across models) leaves every conclusion unchanged: exercise vs.\ reference 23/23, role word 23/23, course path 23/23, file state 21 up and 2 down.

\begin{table}[h]
\centering
\small
\setlength{\tabcolsep}{3pt}
\resizebox{\columnwidth}{!}{%
\begin{tabular}{llrrrr}
\toprule
Mode & Model & none & stub nb. & fin.\ nb. & ex.$-$ref. \\
\midrule
prefill & \model{Coder-1.5B-IT} & 0.6 & 37.0 & 0.4 & $+2.1^*$ \\
 & Coder-7B-IT & 0.0 & 31.9 & 0.6 & $+0.1$ \\
 & Qwen3-8B-IT & 0.0 & 4.3 & 0.0 & $0.0$ \\
free & all three & 0.0 & 0.0 & 0.0 & $\le +0.5$ \\
\bottomrule
\end{tabular}%
}
\caption{\textbf{Chat template.} First-line deferral (\%) on HumanEval. Prefill: the assistant turn starts with the code prompt. Free: the model writes its whole answer. Stub minus finished is significant in prefill on all three models (McNemar).}
\label{tab:chat}
\end{table}

\section{Fill-in-the-middle}
\label{app:fim}
\paragraph{Setup.}
The hole is the function body.
The prefix is a cue plus the HumanEval prompt (as in Section~\ref{sec:frame}); the suffix is the text that follows the function, starting with a blank line.
Each model uses its own sentinel tokens in prefix--suffix--middle order, and we generate greedily until the first non-empty line of the middle; an empty middle counts as deferral.
A gate runs first: with an empty prefix and suffix, the middle must start with an indented body line on at least $80\%$ of HumanEval tasks.
Qwen2.5-Coder-1.5B, -3B, -7B ($0.99$--$1.00$) and StarCoder2-3B ($1.00$) pass; DeepSeek-Coder-1.3B ($0.79$) and CodeGemma-2B ($0.35$) fail and are not reported.
StarCoder2-3B skips the body in left-to-right mode (Section~\ref{sec:setup}), so it is shown but not counted with the Qwen models; it also starts at $36.0\%$ deferral with an empty suffix.
Cues above the hole are the finished or stubbed neighbours and header strings of Section~\ref{sec:setup}.
Cues below the hole are a finished or stubbed function (\ph{} + \code{pass} for the stub) or a doctest \code{\_\_main\_\_} block.
Below the hole the suffix is the next of the three neighbours from Section~\ref{sec:setup} (\code{count\_vowels}, \code{merge\_intervals}, \code{is\_palindrome}), each used once and never the same function as the one above.
Tests are at task level ($n{=}164$, sign-flip permutation, uncorrected): stub below vs.\ none and exercise vs.\ reference both give $p{=}2\times10^{-4}$ on each of the four models.

\begin{table}[t]
\centering
\small
\setlength{\tabcolsep}{3pt}
\resizebox{\columnwidth}{!}{%
\begin{tabular}{lrrrrrrr}
\toprule
 & \multicolumn{3}{c}{Below the hole} & \multicolumn{2}{c}{Above the hole} & \multicolumn{2}{c}{Header} \\
\cmidrule(lr){2-4}\cmidrule(lr){5-6}\cmidrule(lr){7-8}
Model & none & stub & finished & stub & finished & ex. & ref. \\
\midrule
Coder-1.5B & 1.2 & 27.4--49.4 & 0.0 & 74.4 & 2.0 & 50.7 & 23.7 \\
Coder-3B & 8.5 & 97.6--98.8 & 0.2 & 94.9 & 1.0 & 47.2 & 6.5 \\
Coder-7B & 1.2 & 62.8--67.1 & 0.0 & 76.4 & 0.6 & 57.9 & 4.8 \\
StarCoder2-3B$^\dagger$ & 36.0 & 61.6--80.5 & 7.5 & 98.6 & 4.5 & 70.1 & 28.9 \\
\bottomrule
\end{tabular}%
}
\caption{\textbf{Deferral in FIM completion} (first-line, \%, HumanEval, $n{=}164$). \emph{none}: empty prefix and suffix. \emph{Below}: empty prefix; stub and finished functions below the hole (stub: range over the three neighbours; finished: unweighted mean). \emph{Above}: stubbed or finished neighbour above the hole, empty suffix (three neighbours). \emph{Header}: exercise or reference headers, empty suffix (mean of five each). $^\dagger$: skips the body in left-to-right mode.}
\label{tab:fim}
\end{table}

\paragraph{First lines.}
Under a stubbed neighbour above, the model often writes \code{pass}, the neighbour's second line, and skips the comment: on the Qwen models the strict first-line copy rate is $15.0$--$45.5\%$ against $74.4$--$94.9\%$ deferral ($97.6\%$ copy and $98.6\%$ deferral on StarCoder2-3B), so we do not claim form copying for FIM.
On \model{Coder-1.5B} the three functions give $27.4\%$ (\code{merge\_intervals}), $44.5\%$ (\code{count\_vowels}), and $49.4\%$ (\code{is\_palindrome}).
The same three span $1.2$\pp{} on 3B, $4.3$\pp{} on \model{Coder-7B}, and $18.9$\pp{} on StarCoder2-3B.

\begin{figure}[t]
\centering
\includegraphics[width=\columnwidth]{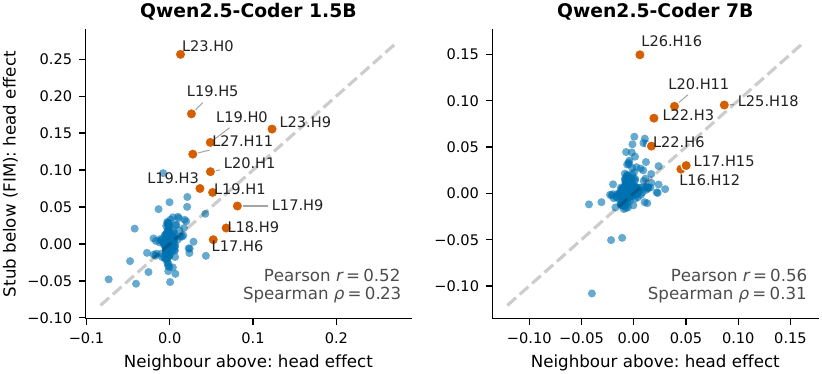}
\caption{\textbf{Head effects of a stub below vs.\ a neighbour above.} Normalised patching effect per attention head: stubbed neighbour above the task (finished into stubbed run, x) and stubbed function below the hole (stubbed into empty suffix, y). Orange: heads in the top five of either axis or the top 15 of both. Correlational; heads are not independent.}
\label{fig:fim_route}
\end{figure}

\paragraph{Head patching.}
For \model{Coder-1.5B} and \model{-7B}, we patch one attention head at a time from a run whose suffix is the stubbed \code{count\_vowels} neighbour into the run with an empty suffix, at the decision positions, and define the normalised effect as in Appendix~\ref{app:route}.
That one function has empty-prefix deferral $44.5\%$ on \model{Coder-1.5B} and $62.8\%$ on \model{Coder-7B}, inside the ranges above.
Figure~\ref{fig:fim_route} compares it with the effect of the neighbour patch of Section~\ref{sec:mech}.
Per-head effects correlate moderately (Pearson $r{=}0.52$ and $0.56$, task-bootstrap 95\% CI $[0.49,0.54]$ and $[0.52,0.57]$; Spearman $\rho{=}0.23$ and $0.31$; $r{=}0.43$ and $0.40$ without the largest $1\%$ of effects).
Five of the top-10 heads overlap on \model{Coder-1.5B} and three on \model{Coder-7B} (about $0.3$ and $0.1$ expected by chance, treating heads as exchangeable).
The largest suffix heads, L23.H0 (\model{Coder-1.5B}) and L26.H16 (\model{Coder-7B}), have prefix effects of $0.01$.
The analysis is correlational, covers two models, and treats heads as independent units; it does not show that the two cues use the same circuit.

\section{Benchmark prompt audit (LiveCodeBench)}
\label{app:lcb}
For stdin problems, the LiveCodeBench completion prompt ends with \code{\# YOUR CODE HERE}.
\model{Coder-1.5B} and \model{DS-1.3B} echo it in $67/67$ allow completions, \model{Coder-7B} in $9/67$, Seed-Coder-8B in $7/67$.
Without the planted line, \model{Coder-1.5B} allow and ban are equal ($35.8\%$), \model{DS-1.3B} is at floor ($0/67$ both), and \model{Coder-7B} gains $+3.0$\pp{} ($2/0$, $p{=}0.50$).
The $+20.9$\pp{} gain from banning comments on LiveCodeBench easy is therefore a prompt-copy effect, and we do not use LiveCodeBench as evidence for the ban effect.

\end{document}